\documentclass[pdflatex,sn-mathphys]{sn-jnl}
\usepackage[T1]{fontenc}
\usepackage{amsmath}
\usepackage{braket}

\begin{document}

\title[Spin wave-driven variable-phase mutual synchronization in spin Hall nano-oscillators]{Spin wave-driven variable-phase mutual synchronization in spin Hall nano-oscillators}
\markboth{}{} 


\author*[1,2,3]{\fnm{Akash} \sur{Kumar}}\email{akash.kumar@physics.gu.se}
\equalcont{These authors contributed equally to this work.} 
\author[1]{\fnm{Avinash K.} \sur{Chaurasiya}}
\equalcont{These authors contributed equally to this work.}
\author[1]{\fnm{Victor H.} \sur{Gonz\'alez}}
\author[1]{\fnm{Nilamani} \sur{Behera}}
\author[1]{\fnm{Roman} \sur{Khymyn}}
\author[1,2,3]{\fnm{Ahmad A.} \sur{Awad}}
\author*[1,2,3]{\fnm{Johan} \sur{\AA kerman}}\email{johan.akerman@physics.gu.se}

\affil[1]{\orgdiv{Department of Physics}, \orgname{University of Gothenburg}, \orgaddress{\street{Fysikgränd 3}, \city{Gothenburg}, \postcode{412 96}, \country{Sweden}}}

\affil[2]{\orgdiv{Research Institute of Electrical Communication}, \orgname{Tohoku University}, \orgaddress{ \city{2-1-1 Katahira, Aoba-ku, Sendai}, \postcode{980-8577} , \country{Japan}}}

\affil[3]{\orgdiv{Center for Science and Innovation in Spintronics}, \orgname{Tohoku University}, \orgaddress{ \city{2-1-1 Katahira, Aoba-ku, Sendai}, \postcode{980-8577} , \country{Japan}}}

\abstract{Spin-orbit torque can drive auto-oscillations of propagating spin wave (PSW) modes in nano-constriction spin Hall nano-oscillators (SHNOs). These modes allow both long-range coupling and the potential of controlling its phase---critical aspect for nano-magnonics, spin wave logic, and Ising machines. Here, we demonstrate PSW-driven variable-phase coupling between two nano-constriction SHNOs and study how their separation and the PSW wave vector impact their mutual synchronization. In addition to ordinary in-phase mutual synchronization, we observe, using both electrical measurements and phase-resolved $\mu-$Brillouin Light Scattering microscopy, mutual synchronization with a phase that can be tuned from 0 to $\pi$ using the drive current or the applied field. Micromagnetic simulations corroborate the experiments and visualize how the PSW patterns in the bridge connecting the two nano-constrictions govern the coupling. These results advance the capabilities of mutually synchronized SHNOs and open up new possibilities for applications in spin wave logic, unconventional computing, and Ising Machines. }

\keywords{Propagating spin waves, mutual synchronization, 
spin Hall nano-oscillators, 
Brillouin light scattering.}

\maketitle

\section*{Introduction}\label{sec1}

The generation, propagation, and control of magnons---the quanta of spin waves---allow long-range transfer~\cite{cornelissen2015long,liu2018long} and processing of digital and analog information~\cite{chumak2014magnon} and form the basis of magnonics~\cite{Chumak2015,pirro2021advances} and spin-wave computing~\cite{Demidov2010,chumak2022advances}. The manipulation of the properties of coherent propagating spin waves (PSWs) in nanoscopic devices, such as their amplitude, phase, propagation direction, and interference patterns, hold great promise for designing magnonic conduits with unique properties~\cite{dieny2020natelectron,barman20212021}. Various emerging applications, including reconfigurable spin-wave logic circuits~\cite{wagner2016magnetic,haldar2016reconfigurable}, unconventional computing~\cite{lee2023task}, and Ising machines~\cite{litvinenko2023spinwave}, rely on these advances. Multiple novel mechanisms have been explored to generate and amplify PSWs~\cite{zhu2023nonlinear,Breitbach2023stimulated}, such as current induced spin-transfer torque (STT)~\cite{bonetti2010experimental,madami2011direct,Houshang2015natnano} and spin-orbit torque (SOT)~\cite{collet2016generation,divinskiy2017advm,evelt2018pra,demidov2020spin,shao2021roadmap}. Nano-constriction spin Hall nano-oscillators (SHNOs) with perpendicular magnetic anisotropy (PMA)~\cite{divinskiy2017apl,fulara2019spin} are a particularly promising approach, 
as they are easy to fabricate~\cite{Demidov2014,durrenfeld2017nanoscale}, CMOS compatible~\cite{zahedinejad2018cmos}, strongly voltage tunable~\cite{fulara2020nt,Zahedinejad2022natmat,choi2022voltage,kumar2022fabrication}, and known for their superior mutual synchronization at various length scales and dimensions~\cite{awad2017natphys,zahedinejad2020nt,kumar2023robust}. 

The mutual synchronization of STT and SOT driven spintronic oscillators is primarily driven by four mechanisms: \emph{i}) dipolar coupling~\cite{Slavin2006,erokhin2014robust}, \emph{ii}) direct exchange~\cite{slavin2009nonlinear}, \emph{iii}) electrical current~\cite{lebrun2017mutual,sharma2021electrically}, and \emph{iv}) PSWs~\cite{slavin2009nonlinear,Houshang2015natnano,chen2009phase,kendziorczyk2014spin, kendziorczyk2016mutual}. Dipolar coupling and direct exchange decay rapidly with distance~\cite{Slavin2006}. While electrical current can couple oscillators over macroscopic distances, it requires magnetic tunnel junction (MTJ) based oscillators with the highest possible magnetoresistance~\cite{lebrun2017mutual,sharma2021electrically}. In contrast, PSWs can drive mutual synchronization over micron distances independent of magnetoresistance~\cite{slavin2009nonlinear,sani2013mutually}. Combining the long-range mutual synchronization of PSWs with the precise control of their frequency, amplitude, and phase will be of great importance for emerging spin wave computing platforms such as spin-wave Ising machines~\cite{houshang2022prappl,litvinenko2023spinwave}.

Here, we report on the experimental observation of variable phase mutual synchronization of nano-constriction SHNOs. The PSWs locally generated by two oscillators separated by $>$200 nm allow radiative locking due to in-phase and out-of-phase coupling of the PSWs. This can be further controlled by the electrical current, the magnetic field, and its orientation. These results are corroborated using state-of-the-art \emph{phase-resolved} micro-focused Brillouin light scattering ($\mu$-BLS) spectroscopy and micromagnetic simulations. The demonstrated control and manipulation of the relative phase of mutually synchronized oscillators at nanoscopic dimensions holds significant promise for various applications such as Ising machines, neuromorphics, and spin-wave computing~\cite{grollier2020nt,chumak2022advances}.

\section*{Results}

\subsection*{Device Fabrication}
Figure~\ref{fig:Schematic}a shows a schematic of the double-nano-constriction SHNOs and the electrical measurement setup. While the in-plane current flow allows for great freedom in designing long chains and large arrays~\cite{awad2017natphys,zahedinejad2020nt,kumar2023robust}, we here focus on double-nano-constrictions to be able to investigate their mutual interaction in isolation and greater detail. To generate PSWs we use W/CoFeB/MgO material stacks (shown in Fig.~\ref{fig:Schematic}b), which offer both interfacial PMA~\cite{ikeda2010natm} and a high spin Hall angle~\cite{fulara2019spin,behera2022energy}. Figure~\ref{fig:Schematic}c shows the scanning electron microscope (SEM) image of the fabricated device (width \textit{w} = 150 nm and separation \textit{d}= 500 nm). We have fabricated devices with varying separation (\textit{d}) from 200 nm to 600 nm. The 200--600 nm separation allows a strong coupling between propagating modes of auto-oscillations in the SHNOs. To compare our results to control samples without PSWs, we also fabricated W/NiFe-based SHNOs without PMA, where the synchronization is driven by dipolar coupling and direct exchange~\cite{awad2017natphys}. The details of material deposition, device fabrication, and electrical measurements are discussed in the Methods section. 

\subsection*{Electrical observation of out-of-phase mutual synchronization}

\begin{figure}
    \centering
    \includegraphics[width=\linewidth]{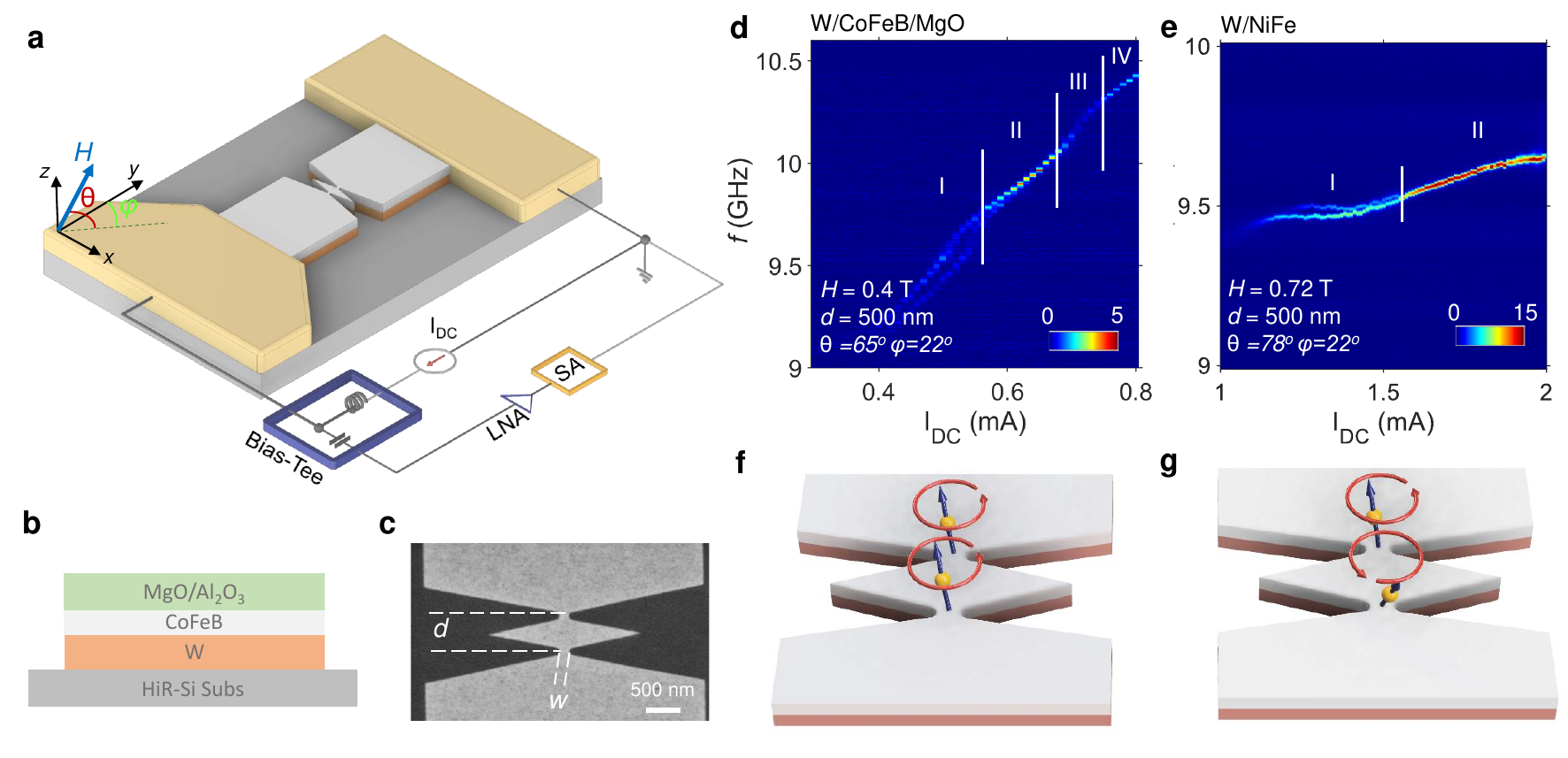}
    \caption{\textbf{Device fabrication and electrical observation :} (a) Schematic of two nano-constriction SHNOs and their connection to the measurement set-up. (b) Configuration of the material stack employed in the fabrication of the W/CoFeB/MgO SHNOs with PSWs. (c) SEM image of the fabricated device with dimensions \textit{w} = 150 nm and \textit{d} = 500 nm. Power spectral density (PSD) \emph{vs.}~applied current (I$_{\rm DC}$) for the nano-constrictions of (d) PMA based W/CoFeB/MgO and (e) in-plane anisotropy based W/NiFe. (f \& g) Illustrations of in-phase and anti-phase mutual synchronization.  }
    \label{fig:Schematic}
\end{figure}

As discussed in detail in~\cite{fulara2019spin}, the PMA raises the auto-oscillating frequency in the nano-constriction above the FMR spin wave gap of the extended magnetic layer, avoiding SW localization and instead promoting the generation of PSWs. The positive and constant non-linearity results in a linearly increasing auto-oscillation frequency as a function of current, corresponding to SWs with increasing wave vector (shorter wave length). Figure~\ref{fig:Schematic}d confirms this quasi-linear current dependence of the auto-oscillation frequency in the W/CoFeB/MgO device (\textit{w} = 150 nm and \textit{d} = 500 nm), which has a threshold current ($I_{\rm th}$) of just below 0.4 mA, an auto-oscillation frequency of about 10 GHz in a 0.4 T field, and about a 15\% increase in frequency when the current is increased to 2$I_{\rm th}$. In comparison, the W/NiFe device (Fig.~\ref{fig:Schematic}e; also two 150 nm wide nano-constrictions separated by 500 nm) without PMA, has a threshold current of about 1.1 mA, needs a field of 0.72 T to reach about the same frequency (no contribution to $H_\mathit{{eff}}$ from the anisotropy field), and starts out with a very weak negative non-linearity changing to a weak positive non-linearity such that the frequency increase is less than 2\% at about 2$I_{\rm th}$. These very different behaviors are consistent with the W/CoFeB/MgO device generating PSWs and the W/NiFe device auto-oscillations being localized. 

The presence/absence of PSWs now leads to different types of mutually synchronized states. The W/CoFeB/MgO nano-constrictions start out in a non-synchronized state (Region I), synchronize from about 0.55 mA to 0.68 mA (Region II), show almost no signal from 0.68 mA to 0.76 mA (Region III), and appear to synchronize again above 0.76 mA (Region IV). In contrast, the W/NiFe nano-constrictions without PSWs only exhibit Regions I \& II. While the high-power signal in Region II results from constructive coherent in-phase interference of the microwave voltage signals from the two mutually synchronized nano-constrictions, corresponding to the state depicted in Fig.~\ref{fig:Schematic}f, Region III represents a new type of behavior, consistent with a possible \emph{anti-phase} mutually synchronized state, depicted in Fig.~\ref{fig:Schematic}g. The absence of a microwave signal could in principle also indicate so-called oscillation death, recently suggested in pairs of interacting MTJ-based STNOs~\cite{wittrock2023non,matveev2023exceptional,perna2024coupling}. However, a faint residue of a single microwave signal can still be observed in Region III, which rules out oscillation death and is instead consistent with an out-of-phase, but not strictly anti-phase, mutually synchronized state. It is noteworthy that Region III and the suggested out-of-phase mutually synchronized state is only observed when PSWs are present. In the W/NiFe device, the mutually synchronized state is robust and shows very high output power, consistent with dipolar coupling and/or direct exchange being responsible for the coupling as they both favor in-phase mutual synchronization.

\subsection*{$\mu$-BLS microscopy of the individual nano-constrictions}

\begin{figure}[b]
    \centering
    \includegraphics[width=\linewidth]{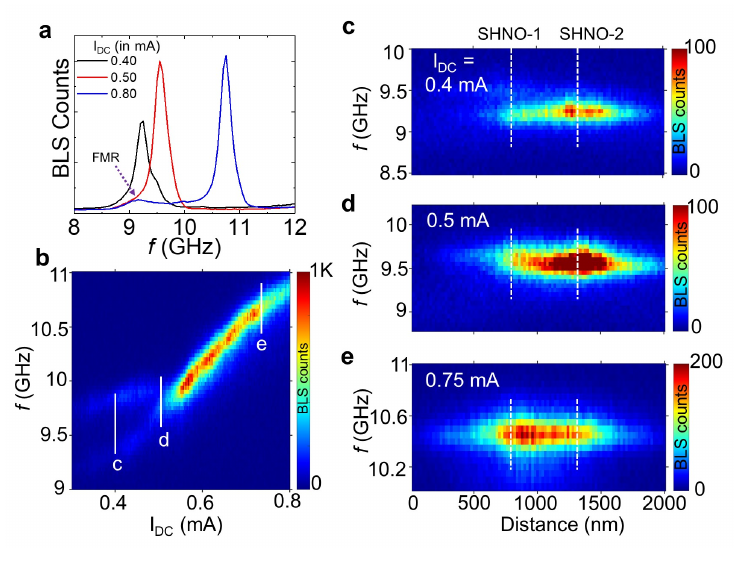}
    \caption{\textbf{Spatial mapping using $\mu$-BLS}: (a) Representative BLS spectra showing FMR as well as the auto-oscillations measured at I$_{\rm DC}=$ 0.40, 0.50, and 0.80 mA. (b) Current-dependent auto-oscillation signal measured using $\mu$-BLS. (c-e) Spin-wave intensity profiles of the double SHNOs along the constrictions, measured at applied current I$_{\rm DC}=$ 0.4, 0.5, and 0.75 mA, respectively. The dotted lines refer to the position of the constrictions.}
    \label{fig:spacial_maps}
\end{figure}

To conclusively rule out oscillation death and directly visualize the auto-oscillations in each nano-constriction, we now turn to results from $\mu$-BLS. We first used conventional $\mu$-BLS microscopy to map out the SW intensity \emph{vs.}~both frequency and spatial coordinates in the double-nano-constrictions. The magnetic field conditions were the same as in the electrical measurements above (details of the measurements can be found in the Methods section). Figure~\ref{fig:spacial_maps}a first shows the spectral content of the SWs at three different currents, as measured on the bridge connecting the two nano-constrictions (same device as in Fig.~\ref{fig:Schematic}d). We here want to highlight how the high-intensity SW auto-oscillations all lie above the weak thermally excited FMR peak at about 9.1 GHz, confirming their propagating nature. Figure~\ref{fig:spacial_maps}b then focuses on the full current-dependent spectral distribution of the auto-oscillations, displaying both important similarities and differences when compared with the electrical data in Fig.~\ref{fig:Schematic}d. At low currents, we observe two faint signals with about the same threshold as in Fig.~\ref{fig:Schematic}d. At about 0.55~mA, the two signals merge and the BLS counts increase strongly and remain high for all higher currents. As in the electrical measurements, the frequency dependence is essentially linear in current, consistent with PSWs above FMR. At low current, the linearity is better for the lower-frequency signal whereas the higher-frequency signal experiences considerable frequency pulling until the two merge.

It is straightforward to identify the behavior below 0.55~mA with two un-synchronized nano-constrictions, and hence identical to Region I in the electrical data. However, above 0.55~mA the situation is different. Whereas it is again straightforward to identify the state above 0.55~mA with two mutually synchronized nano-constrictions, \emph{i.e.}~with Region II, there is no sign of any transition at 0.68~mA into a Region III with strongly decreasing BLS counts. Instead, the BLS counts remain essentially constant across 0.68~mA and continue to display all characteristics of a mutually synchronized state with high SW intensity. This rules out the possibility of oscillation death being the reason for the very weak electrical signal in Region III. 

To gain further insight into the auto-oscillation modes, we present in Fig.~\ref{fig:spacial_maps}c-f hybrid frequency-spatial BLS maps for a few selected I$_{\rm DC}$ along a line through the double-nano-constrictions. At I$_{\rm DC}=$ 0.4 mA, the spatial maps indicate an unsynchronized state, with SHNO-1 having a higher frequency but lower counts than SHNO-2 (there is substantial leakage of the SHNO-2 signal into the SHNO-1 region due to the ~300 nm laser spot size; this should not be interpreted as SHNO-1 auto-oscillating on this frequency). At I$_{\rm DC}=$ 0.5~mA, the two oscillators are close to, but not yet, synchronizing. SHNO-1 now has higher counts and its frequency has been pulled closer to that of SHNO-2. The BLS map remains asymmetrical about its central frequency, indicating that the two regions are not yet mutually synchronized. 
However, at I$_{\rm DC}=$ 0.75~mA, \emph{i.e.}~well inside Region III, the BLS map is symmetrical across its central frequency and both SHNOs show higher counts, indicative of a mutually synchronized high-intensity state. Again, this rules out oscillation death being a possibility and corroborates out-of-phase mutual synchronization as the more likely explanation. To directly measure the internal relative phase of the mutually synchronized state, we will therefore resort to \emph{phase-resolved} $\mu$-BLS microscopy.

\subsection*{Direct observation of out-of-phase mutual synchronization using phase-resolved $\mu$-BLS microscopy}

\begin{figure}[t]
    \centering
    \includegraphics[width=\linewidth]{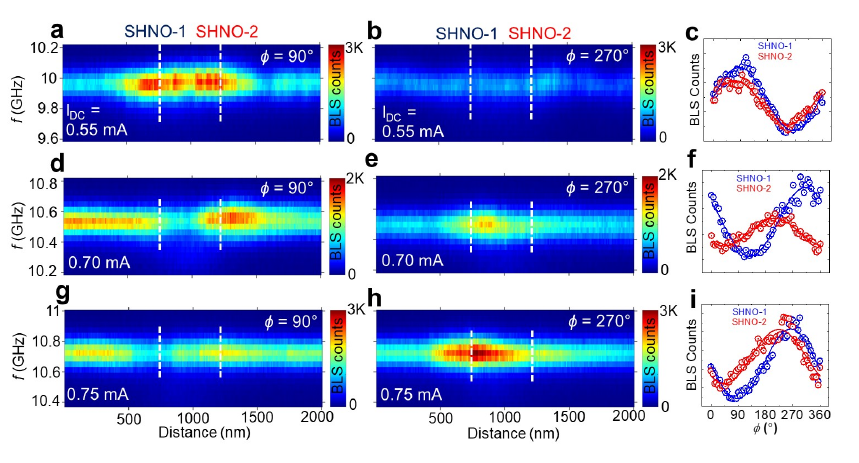}
    \caption{\textbf{Phase-resolved $\mu$-BLS measurements:} \emph{Phase-resolved} spin-wave intensity maps of the double SHNOs measured at I$_{\rm DC}=$ 0.55 (a, b), 0.70 (d, e) and 0.75 mA (g, h), with two different phase settings ($\phi$) separated by 180$^\circ$. The last column (c, f, i) shows the BLS counts as a function of $\phi$ measured at the centre of each nano-constriction. The symbols are the measured counts at an injection of $P_\mathit{IL}=$ -10 dBm; the solid lines are sinusoidal fits.}
    \label{fig:phase_maps}
\end{figure}

As described in more detail in Methods section, \emph{phase-resolved} $\mu$-BLS microscopy can determine the phase of the detected SWs \emph{w.r.t.}~a reference signal. This is usually done by exciting the SWs directly with the reference signal, \emph{e.g.}~fed to an antenna~\cite{serga2006phase}. However, in SHNOs the SWs are generated intrinsically by auto-oscillations and to study their phase, one must first injection lock the SHNO to the reference~\cite{houshang2022prappl}. Strictly speaking, \emph{phase-resolved} $\mu$-BLS microscopy does not study the free-running nano-constriction pair, only the corresponding injection-locked system, which may or may not be similar to the free-running mutually synchronized state. To keep the perturbation to a minimum, in the hope of extracting an oscillation phase not too different from the free-running case, we keep the power of the injection-locking signal to a minimum, still ensuring a stable injection-locked state. 

Figure~\ref{fig:phase_maps} shows the \emph{phase-resolved} $\mu$-BLS results from the W/CoFeB/MgO device at three different current levels: I$_{\rm DC}=$~0.55 mA, corresponding to Region II, and I$_{\rm DC}=$~0.70 \& 0.75 mA, corresponding to two different points in Region III. At all times, the device is injection-locked with minimal power $P_\mathit{{IL}}=$~--10 dBm and $f_\mathit{{IL}}=$~$f_\mathit{{SHNO}}$. Figure~\ref{fig:phase_maps}a shows a hybrid frequency-spatial map of the phase-BLS counts as a function of frequency and position along the line connecting the two nano-constrictions. Here, the phase angle \emph{w.r.t.}~the reference is set to $\phi=$ 90$^\circ$ (controlled by an electrical phase shifter). Figure~\ref{fig:phase_maps}b shows the corresponding counts when the phase shifter is rotated to $\phi=$ 270$^\circ$. It is evident from these two plots that the two nano-constrictions are in phase with each other and contribute about equal counts to the BLS intensity. Figure~\ref{fig:phase_maps}c shows the full phase-dependent BLS counts extracted from the locations of the two nano-constrictions (dashed white lines) when $\phi$ is varied from 0$^\circ$ to 360$^\circ$. Sinusoidal fits to the experimental data yield a small relative phase difference of $\Delta \phi=$ 17$\pm3$$^\circ$ between the two SHNOs. The phase-resolved $\mu$-BLS results hence corroborate the conclusion from the electrical measurements that the two nano-constrictions largely auto-oscillate in phase.

However, the situation is dramatically different in Region III. Figures~\ref{fig:phase_maps}d--i show the corresponding phase-dependent results at I$_{\rm DC}=$~0.70 \& 0.75 mA. The two nano-constrictions show very different behavior in the BLS maps and when the full phase-dependent counts are fitted, we extract very large relative phases of $\Delta \phi=$ 100$\pm5$$^\circ$ at 0.70 mA and $\Delta \phi=$ 51$\pm5$$^\circ$ at 0.75 mA. As already indicated by the electrical data, Region III is hence conclusively characterized by out-of-phase mutual synchronization, which explains the almost vanishing electrical microwave signal in this region.

\subsection*{Micromagnetic simulations}
To corroborate our experimental findings we carry out micromagnetic simulations using mumax$^3$~\cite{Vansteenkiste2014} to see whether we can reproduce and further understand the experimentally observed behavior. The simulated device is identical to the W/CoFeB/MgO double-SHNO with two 150 nm wide nano-constrictions separated by 500 nm. The magneto-dynamical parameters for the simulation were extracted from experimental data obtained through spin-transfer ferromagnetic resonance (STFMR) measurements on W/CoFeB/MgO microstrip devices~\cite{behera2022energy,kumar2023robust}. A detailed description of the simulations is presented in the Methods section. 

\begin{figure}[h]
    \centering
    \includegraphics[width=0.8\linewidth]{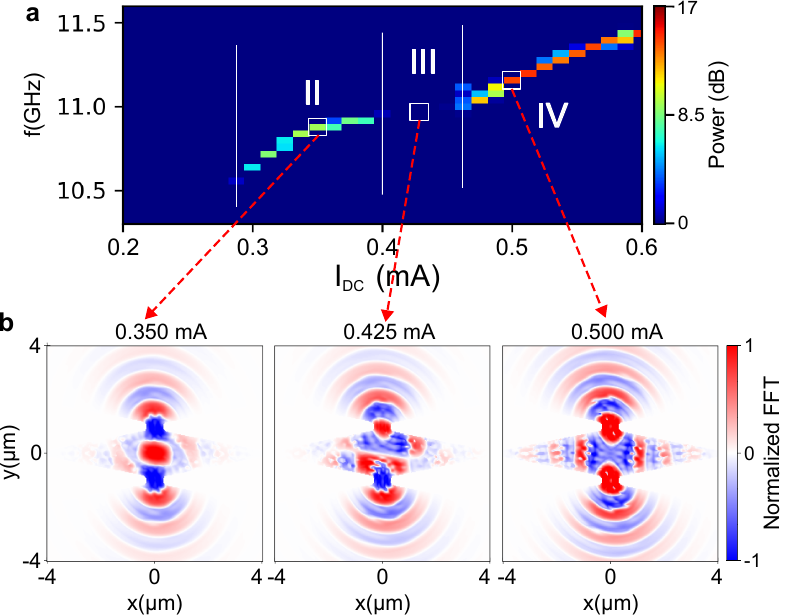}
    \caption{\textbf{Micromagnetic simulations:} (a) Simulated PSD \emph{vs.}~I$_{\rm DC}$ for two 150 nm wide nano-constriction SHNOs separated by 500 nm, reproducing the three mutually synchronized regions (II--IV) observed in the electrical measurements. 
    (b) Complex auto-oscillation mode profiles at currents in each region. We observe that the shape of the resonant modes in the bridge connecting the constriction influences their stationary phase convergence.}
    \label{fig:PSD-micromag-sims}
\end{figure}

Figure~\ref{fig:PSD-micromag-sims}a shows the simulated PSD plotted against the applied direct current, which reproduces the experimental results remarkably well with minor expected differences: \emph{i}) the threshold current in the micromagnetic simulations ($T=$ 0 K) is slightly lower than in the room-temperature experiments, and \emph{ii}) the two identical nano-constrictions synchronize already at threshold as they auto-oscillate on exactly the same frequency, \emph{i.e.}~there is no Region I. Except for this region, we can now readily identify the same regions as in Fig.~1d above: II) a high-power in-phase mutually synchronized state, III) the disappearance of the microwave signal, corresponding to an anti-phase mutually synchronized state, and IV) the reappearance of a strong microwave signal, again corresponding to in-phase mutual synchronization. Figure~\ref{fig:PSD-micromag-sims}b shows spatial maps of the complex Fourier transform at the corresponding auto-oscillating frequencies in Regions II--IV, where the phase of the SWs confirms the in-phase mutual synchronization in Regions II \& IV, and the anti-phase mutual synchronization in the middle of Region III. Furthermore, we can discard oscillator death as a possible explanation for this behavior, as we can study the oscillators individually as shown in Fig. S6a in the supplementary information.

\subsection*{Current-controlled variable-phase mutual synchronization}

\begin{figure}[b]
    \centering
    \includegraphics[width=0.57\linewidth]{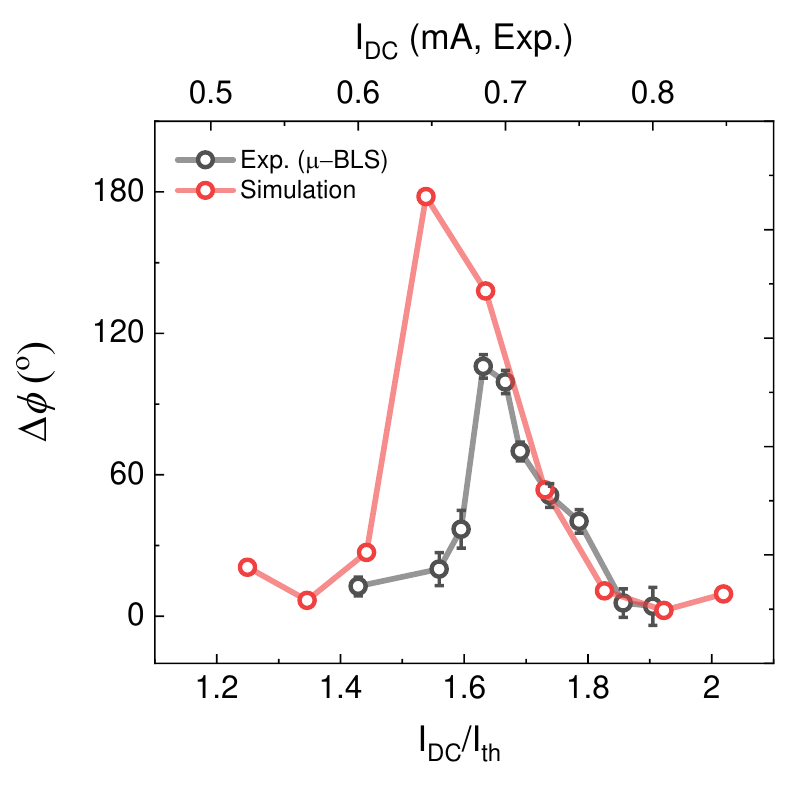}
    \caption{\textbf{Evolution of the relative phase:} The variable phase difference $\Delta$$\phi$ between two mutually synchronized SHNOs as a function of criticality (I$_{\rm DC}$/I$_{\rm th}$) obtained from experiments ($\mu$-BLS) and simulations (micromagnetics). The top \textit{x}-axis shows I$_{\rm DC}$ (in mA) for the experimental data.}  
    \label{fig:variable_phase}
\end{figure}

We finally demonstrate how the drive current can be used to continuously tune the internal phase of the mutually synchronized state, and hence the phase of the coupling between two SHNOs. Figure~\ref{fig:variable_phase} shows the current dependence of the internal relative phase difference, $\Delta \phi$, of the mutually synchronized state, as extracted from \emph{phase-resolved} $\mu$-BLS measurements on the W/CoFeB/MgO device. The experimental data shows how the internal phase is essentially zero at low current (0.60 mA), rapidly increases to a maximum of 106$^\circ$ at intermediate current (0.685~mA), and then, more gradually, decreases towards zero at the highest current (0.80 mA). The phase difference is hence possible to tune continuously with current. The micromagnetic simulations reproduce this behavior remarkably well, albeit with about a 1.5 times greater phase difference at the peak. We ascribe this difference to the aforementioned limitation of \emph{phase-resolved} $\mu$-BLS microscopy, as it requires the device to be injection locked in order to extract the phase. While the micromagnetic simulations can extract the true $\Delta \phi$ of the simulated SHNOs, the injection locking signal will reduce the experimental $\Delta \phi$ when it interacts with the oscillators, trying to align both their phases with its own. This can be demonstrated experimentally by increasing the strength of the injection locking signal beyond the minimum value for injection locking (Fig. S5 in the supplementary information). A plot of $\Delta \phi$ \emph{vs.}~$P_\mathit{IL}$ (Fig.~S5) shows how the extracted relative phase decreases from 100$^\circ$ to about 60$^\circ$ with increasing injected power. It is reasonable to expect this trend to continue if yet lower power could have been used. The intrinsic, unperturbed value is hence out of reach. Even so, the overall trends of the experiment and the simulation in Fig.~\ref{fig:variable_phase} agree very well.

\subsection*{Varying the nano-constriction separation}

In addition to the detailed results above, we explored a large number of additional devices under different measurement conditions to confirm the robustness, repeatability, and control of the variable-phase mutual synchronization. For example, changing the nano-constriction separation should have an immediate impact on the phenomenon as different wave vectors would match different separations. Detailed results of these extensive studies are provided in the supplementary information, sections S1-S4. In Fig.~S1, we compare the current dependent PSD of W/NiFe and W/CoFeB/MgO nano-constriction pairs with 300 nm, 400 nm, and 500 nm separation, respectively. As in Fig.~\ref{fig:Schematic}e, the W/NiFe pairs consistently show only Region I \& II, with Region II starting at increasingly higher currents the greater the separation. This is consistent with the coupling mechanism being dipolar and/or direct exchange, essentially identical to earlier experimental reports~\cite{awad2017natphys}. In contrast, the three W/CoFeB/MgO devices all show regions of disappearing microwave signal, indicating out-of-phase mutual synchronization. 

To investigate whether the location of Region III depends systematically on nano-constriction separation ($d$), we fabricated a large set of nano-constriction pairs with finer steps in $d$. PSD measurements of 11 such devices with $d$ ranging from 200 to 560 nm are shown in Fig.~S2. While all devices show regions of disappearing microwave signal, the location of these regions does not appear to be particularly systematic. 

\subsection*{Varying the applied magnetic field}

Fig.~S3a-l depicts the PSD \emph{vs.}~direct current, measured for field strengths ranging from 0.35 T to 0.46 T with a step size of 0.01 T for W/CoFeB/MgO SHNOs double nano-constrictions separated by 420 nm. As expected, both the auto-oscillation frequency and threshold current increase quasi-linearly with field strength. The location of Region III also depends on the field strength in a systematic manner. In Fig.~S3m we summarize the results by extracting the central frequency of the auto-oscillations only when the PSD is above a certain value. This approximately captures the beginning and the end of Region III for all field strengths. Much in the same way as the threshold current increases with field strength, the location of Region III shifts quasi-linearly to higher currents with higher fields. To a first approximation, the current difference between the location of Region III and the threshold current stays constant with increasing field. This is consistent with the wave vector of the PSWs being independent of the external field strength, but directly dependent on the criticality I$_\mathit{DC}$/I$_\mathit{th}$ (see Discussion, below).

Fig.~S4 shows the PSD \emph{vs.}~direct current as a function of the out-of-plane angle ($\theta=$ 55$^\circ -$ 68$^\circ$) for the same double nano-constriction. The dependence on $\theta$ is considerably more complex than on field strength, with two regions of signal extinction appearing at lower angles, which appear to merge at higher angles.

\section*{Discussion}

The strong PMA of the W/CoFeB/MgO material stack counteracts the shape anisotropy and negative non-linearity of the thin film geometry, and, with the additional help from a moderate applied field, pulls the magnetization out-of-plane and turns the non-linearity positive; this leads to magnon-magnon repulsion and the excitation of PSWs~\cite{fulara2019spin}. Thus, the qualitative behavior of the mutual synchronization of SHNOs should be approximately described using the ordinary SW dispersion of an out-of-plane magnetized film~\cite{Houshang2018natcomm}: $f=f_{\text{FMR}}+\frac{\gamma}{2\pi} D k^2$, where $f_{\text{FMR}}$ is the frequency of the ferromagnetic resonance and $D \simeq 2 A_{ex}/M_s$ the dispersion coefficient. Our micromagnetic simulations show that the PSW has a substantial wavevector already at the auto-oscillation threshold, mainly defined by the geometry of the constriction. This is largely identical to the original description of PSWs in nano-contact STNOs \cite{Slonczewski1999jmmm}, where the wavevector at threshold is given by the nano-contact diameter. With increasing criticality (current increasing beyond the threshold) the wavevector further increases. Considering the frequency at the start of Region II: $f_{\text{II}}=9.75$ GHz and III: $f_{\text{III}}=10.1$ GHz (Fig.\ref{fig:Schematic}. d) and taking $f_{\text{FMR}}=9.3$ GHz (Fig.\ref{fig:spacial_maps} b) one gets the wavelengths $\lambda_{\text{II}}=304$ nm and $\lambda_{\text{III}}=228$ nm. This is in good agreement with the results of the simulations for PSWs outside of the constriction region (see Fig. \ref{fig:PSD-micromag-sims}). Interestingly, since the applied magnetic field contributes mainly to the first term of the dispersion law, \emph{i.e.}~to $f_{\text{FMR}}$, and anti-phase locking occurs at the same wavevector for different field values, Region III moves in parallel with the threshold current when the field strength is varied (Supplementary Fig.~S3).

We note that the obtained wavelengths do not coincide with the distance between SHNOs since they are related to PSWs \emph{outside} the nano-constriction and bridge region. Moreover, our experiments \emph{vs.}~nano-constriction separation ($d$) did not show a clear dependence of the position of Region III (Fig.~S2). The micromagnetic simulations provide additional insight into this behavior as they reveal the crucial importance of the SW patterns inside the rhombic bridge connecting the two nano-constrictions. The complex profiles of the SWs within the rhombic bridge are drastically different from the freely propagating SWs outside the SHNOs and highlight the importance of the particular SW modes of this area (see Fig.~\ref{fig:PSD-micromag-sims}b). As a result, the distance between two in-phase oscillating points here is very different from the wavelength of a free SW. The dependence on the details of the SW modes of the bridge also explains the complexity of how Region III depends on the out-of-plane angle. The angle has a strong impact on the non-linearity, with non-trivial consequences for the particular SW modes that will dominate in the bridge.

The sensitivity to the details of the bridge also explains the lack of systematics when we varied $d$ in Fig.~S2. Conversely, this sensitivity should allow for sensitive control of the phase of the mutual synchronization, both through the shape and dimensions of the bridge, and, more interestingly, via voltage-control of the PMA~\cite{fulara2020nt} in the bridge region. As the PSWs fill up the bridge region, it might be sufficient to fabricate voltage gates on the two sides of the bridge to avoid any detrimental processing damage in the central region between the two nano-constrictions. While we have focused on rhombic bridges in this study, there is great freedom in future bridge designs with or without voltage gates placed at different locations. This will lead to very rich variable-phase phenomena in the coupling between adjacent nano-oscillators, with direct application in neuromorphic computation and Ising Machines.

\section*{Methods}\label{sec13}
\subsection*{Spin Hall nano-oscillator fabrication}
Thin film stacks of W (5 nm)/CoFeB (1.4 nm)/MgO (2nm)/Al$_2$O$_3$ (4 nm) and W (5 nm)/NiFe (3 nm)/Al$_2$O$_3$ (4 nm) were prepared on a high resistive Si substrate (HiR-Si, $\rho>$20,000 $\Omega-cm$) using DC/RF magnetron sputtering (AJA Orion 8) at room temperature. The W/CoFeB/MgO thin films were annealed post deposition for an hour at 300$^\circ$ under ultra high vacuum to induced interfacial PMA. The SHNO devices with 150 nm widths were fabricated using e-beam lithography (Raith EBPG 5200) followed by Ar-ion milling.~\cite{kumar2022fabrication} For the mutual synchronization experiments, double-nano-constrictions were fabricated with varying separation (200-600 nm). The Ground-Signal-Ground (G-S-G) contact pads were fabricated in a subsequent step using mask-less UV lithography (Heidelberg Instruments MLA 150) and lift-off technique, deposition of Cu (800 nm)/Pt (20 nm) for contact pads are performed using DC magnetron sputtering.

\subsection*{Electrical measurements}
The electrical measurements for the characterization of free-running properties of SHNO devices were performed using a custom-designed G-S-G pico-probe setup (150 $\mu$m pitch from GGB industries) between the electro-magnet poles. The sample stage has the functionality of motorized out-of-plane rotation, for applying out-of-plane magnetic field. A direct current was supplied to the SHNO devices using a constant current source (KE 6221). A magnetic field of 0.4--0.8 T was applied at 65$^\circ$ out-of-plane and 22$^\circ$ in-plane angle to achieve propagating spin-wave modes in CoFeB thin films~\cite{dvornik2018pra,fulara2019spin}. The generated microwave auto-oscillations were amplified using a low-noise amplifier (LNA, 32 dBm from BnZ Technologies) and observed using a spectrum analyzer (SA, R$\&$S FSV) with resolution bandwidth (RBW) of 1 MHz. All measurements were performed at room temperature.

\subsection*{Phase-resolved $\mu-$BLS measurements}
The magneto-optical measurements were performed using the micro-focused Brillouin light scattering technique. A monochromatic continuous wave laser (wavelength = 532 nm) was focused on the nano-constriction region by 100$\times$ microscope objective having a large numerical aperture (NA = 0.75) down to 300 nm diffraction limited spot diameter. The magnetic field conditions were kept almost identical to the one used in the electrical measurement. To get the \emph{phase-resolved} information, some fundamental modification in the setup was made. Here, we modulated the phase of the incoming light using an electro-optical modulator at the same frequency as the injection signal applied to the double nano-constriction SHNOs. The basic principle relies on the interference of the elastically scattered light (controllable phase via electrical phase-shifter) and inelastically scattered light from the oscillators carrying phase information ~\cite{serga2006phase,sebastian2015}. The resulting BLS signal is hence the difference between the light (known) and the oscillator (unknown). The resultant signal was analyzed with a Sandercock-type six-pass Tandem Fabry-Perot interferometer TFP-1 (from JRS Scientific Instruments). A three-axis nanometer-resolution stage, along with an active stabilization protocol provided by THATec Innovation GmbH, was employed to get precise long-term spatial stability during the measurement. All the measurements were performed at room temperature.  

\subsection*{Micromagnetic Simulations}

The micromagnetic simulations of the ferromagnetic (FM) layer were performed using the GPU-accelerated program mumax3~\cite{Vansteenkiste2014} with volumetric current density and Oersted field generated using COMSOL multiphysics 6.1. The e-beam lithography schematics used for sample fabrications were imported directly into COMSOL and a bilayer device of W (5~nm)/CoFeB(1.4~nm) was simulated using the \textit{Magnetic and electric fields (mef)} package, with the materials' electrical properties taken directly from our measurements.

The 4~$\mu$m $\times$ 4~$\mu$m $\times$ 1.4~nm double nano-constriction SNHO geometry obtained from the fabrication schematics was discretized into 512 $\times$ 512$\times$ 1 cells. The material parameters used in the simulations were either measured directly from the samples using FMR (saturation $M_S$= 1050 kA/m, gyromagnetic ratio $\gamma/2\pi$ = 29.1 GHz/T, Gilbert damping constant $\alpha$ = 0.025 and PMA field of $K_u$= 645~kJ/m$^3$) or taken from the literature (A$_{ex}$ = 19$\times$10$^{-12}$~\cite{Sato2012ExchangeStiffness}). This system was excited with DC biasing currents between 100-600 $\mu$A. The OOP spin-polarized currents from the W layer were calculated using the spin Hall angle $\theta_{SH}$= 0.6, measured from microstrip devices~\cite{behera2022energy}. The torque generated by this current was calculated at each cell and added using a fixed layer at the bottom of the CoFeB film. The magnetic dynamics of the FM film were simulated by integrating the Landau-Lifshitz-Gilbert-Slonczewski equation over 30 ns. It was found that the relative phase between constrictions settled after 15 ns for all evaluated currents. 

The power spectral densities were calculated using a complex FFT of the time evolution of the average magnetization of the whole sample and each of the constrictions. The mode profiles of the device were obtained using complex point-wise FFT on the full magnetization maps. The phase difference between the nano-constrictions was extracted from line scans of the AO mode profiles as detailed in Fig.S6b in the supplementary information. 

\section*{Declarations}

\bmhead{Funding}
This work was partially supported by the Horizon 2020 research and innovation program No. 835068 "TOPSPIN". This work was also partially supported by the Swedish Research Council (VR Grant No. 2016-05980).

\bmhead{Conflict of interest}
The authors declare no competing interests.

\bmhead{Availability of data and materials}
The data supporting the findings of this study are available within the paper and other findings of this study are available from the corresponding author upon reasonable request.

\bmhead{Authors' contributions} 
AK and JÅ initiated the idea. AK, AKC, AAA, and JÅ designed the experiments. AK and NB optimized the thin film stacks. AK fabricated devices and performed electrical measurements. AKC performed the BLS measurements with input from AAA. VHG performed the micromagnetic simulation with input from RK. Both AK and AKC analyzed the data; J.Å. managed the project; all co-authors contributed to the manuscript, the discussion, and the analysis of the results.

\bibliography{references}

\end{document}